\documentclass[preprint]{aastex}

\title{Remote Operations and Nightly Automation of The Red Buttes Observatory}
\author{David H. Kasper\altaffilmark{1},
	Tyler G. Ellis\altaffilmark{1},
	Rex R. Yeigh\altaffilmark{1},
	Henry A. Kobulnicky\altaffilmark{1},
	Hannah Jang-Condell\altaffilmark{1},
	Mark Kelley\altaffilmark{2},
	Gerald J. Bucher\altaffilmark{1},
	James S. Weger\altaffilmark{1}
	}
\email{dkasper@uwyo.edu, tellis8@uwyo.edu}
\altaffiltext{1}{Dept. of Physics \& Astronomy, University of Wyoming, Laramie, WY 82070, USA}
\altaffiltext{2}{DFM Engineering, Inc., 1035 Delaware Ave, Longmont, CO 80501}


\usepackage{graphicx,amssymb,amsmath,float,natbib,listings}
\begin{document}

\begin{abstract}
We have implemented upgrades to the University of Wyoming's Red Buttes Observatory (RBO) to allow remote and autonomous operations using the 0.6 m telescope. Detailed descriptions of hardware and software components provide sufficient information to guide upgrading similarly designed telescopes. We also give a thorough description of the automated and remote operation modes with intent to inform the construction of routines elsewhere. Because the upgrades were largely driven by the intent to perform exoplanet transit photometry, we discuss how this science informed the automation process. A sample exoplanet transit observation serves to demonstrate RBO's capability to perform precision photometry. The successful upgrades have equipped a legacy observatory for a new generation of automated and rapid-response observations.
\end{abstract}
\maketitle
\section{Introduction}
The Red Buttes Observatory (RBO) was constructed in 1994 as the second research observatory at the University of Wyoming (UW). Located at 2,240 m elevation, the observatory lies 15 km south of Laramie, Wyoming at 41.3167\degr\ N, 105.5833\degr\ W. Its 0.6 m \textit{f}/8.43  Ritchey-Chr\'{e}tien Cassegrain telescope was constructed and installed by the Colorado-based company DFM Engineering, Inc (DFM). RBO began operation with the mission of enriching the undergraduate astronomy curriculum with practical experiences in observation. Successful research projects have included prototyping new instrumentation, measuring eclipsing binary distances to star forming regions, and followup confirmation of $\gamma$-ray bursts \citep{Rodgers2006, Monson2009, Kiminki2015}. The observatory was operated for two decades in the classical fashion with on-site observers.

We initiated a series of upgrades in 2014 to the telescope control system and observatory infrastructure to enable remote and autonomous operation. The project's goal was to increase the scientific productivity of the observatory by enabling nightly observations without on-site real-time human interaction. A remote-enabled observatory mitigates logistical issues of Wyoming weather and department resources, permitting increased year-round observing. Renovations have been completed which have enabled remote and autonomous operation of the observatory.

We found in the course of the upgrades, that there existed little literature to guide the conversion of small legacy observatories to autonomous operation. Our goal in this publication is to describe the RBO upgrade process in sufficient detail such that observatories of similar vintage can enact like upgrades. DFM has installed 24 telescopes which have similar control systems to RBO. Section 2 details the hardware and software implemented in the automation process. We also provide details regarding the in-house scripts required to interface the observatory devices in this section. Section 3 describes the event sequence during a typical remote or automated night. Section 4 describes the science driver for RBO, how it informed the automation process, and a sample dataset of a transiting exoplanet to demonstrate current photometric capabilities.

\section{Upgrade Implementation}
The upgrades at RBO can be categorized as either hardware- or software-based. In the first subsection, we detail the hardware components necessary for remote and autonomous operation. In the second subsection, we describe software and additional scripts used to interface components of the observatory. Emergency contingencies are included with the descriptions of each relevant component.
\begin{figure}
	\begin{center}
		\includegraphics[width=.7\linewidth]{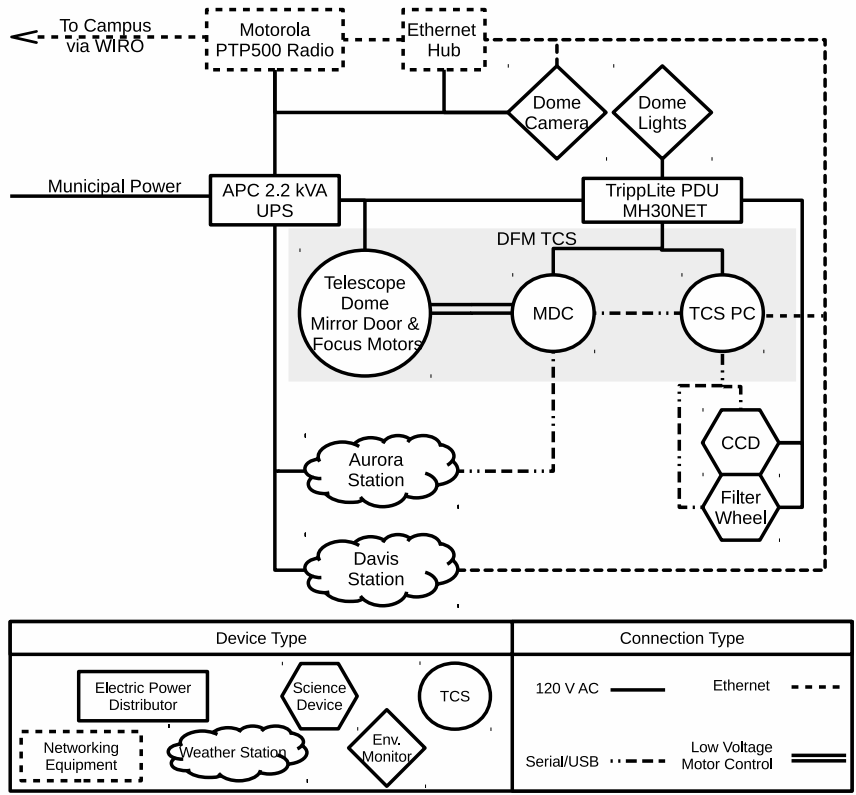}
		\caption{\label{fig:rbodiagram}A schematic of all necessary remote/automated observation components at RBO. The grey shaded area encapsulates the DFM TCS. The key indicates the meaning of the different line styles and symbol shapes.}
	\end{center}
\end{figure}
	\subsection{Operation Hardware}
	In the Summer of 2014, DFM designed and implemented a next-generation telescope control system (TCS), replacing the original 1994 DFM proprietary TCS at RBO. Figure \ref{fig:rbodiagram} is a schematic overview of the hardware components, power, and communications infrastructure of the new DFM control system and the observatory at large utilized in remote/automated operations. The gray shaded area encapsulates the TCS. Single solid lines denote 120 V AC power, dash-dot lines represent serial or USB connections, dashed lines indicate ethernet and network connections, and solid double lines indicates low-voltage motor control or TTL connections. The rest of this subsection will describe each of the main components in the figure.

	The new TCS is a hybrid, consisting of DFM proprietary code running on a 32-bit Windows 7 operating system computer (TCS PC) and an external, commercial, solid state motion controller (Galil DMC 4183) with embedded code. The motion controller provides the real-time control interface for commanding the telescope, dome, mirror doors, and focus motor speeds. The motion controller also mediates hand paddle inputs and updates the states of dome/tracking switches.\footnote{Switches included on the front panel of legacy DFM control systems include those to turn on telescope tracking, dome rotation, and mirror doors. These switches, formerly activated by an on-site user, now remain in the ``on'' position for remote and automated operation.} The motion controller and additional custom circuits --- providing linear servo power, electronic surge suppression, horizon interlock circuitry, solid state relays for dome control, and long-cable signal conditioning --- reside in a rack-mounted Motor Driver Chassis (MDC) adjacent to the TCS PC.
	
	The TCS provides control of the telescope, focus, dome rotation, dome upper and lower shutters, and mirror door covers. Communication between the TCS PC and motion controller occurs over TCP/IP on a dedicated local area network using a dedicated network interface card. A key feature of the TCS, implemented in the updated code, is its inclusion of ASCOM (Astronomy Common Object Model) compliant drivers that allow users and third party software to gather telescope, dome, and focus status and issue commands through the established ASCOM standard protocols. These protocols have been available since 1999 and are now widely deployed on small telescopes.\footnote{http://ascom-standards.org}  
	
	Other hardware upgrades included replacing the original	friction-driven quadrature encoders on each telescope axis with Heidenhain ROC 425 25-bit on-axis absolute encoders. The on-axis encoders provide the TCS with absolute telescope position without requiring initialization. The dome position encoder, driven by a gear on the dome track, was augmented by the addition of a ``dome home'' fiducial switch at azimuth 270\degr\ to ensure absolute positioning. Together these upgrades ensure accurate knowledge of telescope and dome position, even in the event of an unexpected shutdown or power failure. A proprietary Heidenhain IK220 PCI card within the TCS PC provides the interface between the ROC425 encoders and the TCS code.
	
	Because there is no line-of-sight path between the UW campus and RBO, internet is provided by a two-segment microwave radio link, with a relay point at the Wyoming Infrared Observatory (WIRO) on Mt. Jelm,  which has a line-of-sight path to both locations. The 34 mi RBO-to-WIRO segment uses by a 25 Mb s$^{-1}$  5 GHz unlicensed radio link, with a Motorola PTP 500 series radio at each end. The 40 mi WIRO-to-UW campus link occurs over a 300 Mb s$^{-1}$ licensed 11 GHz radio link with a pair of Motorola PTP 800 series radios and 2 ft diameter antennas. The two segments converge at WIRO within the local area network anchored by an ethernet switch. RBO's internet bandwidth is sufficient for all needs of a modern observatory (i.e., simultaneous data transfer of 4 MB images or larger, remote desktop connection, and audio/video feed). In the event of a network outage, a weather parsing script triggers a dome closure to ensure observatory safety. The details of this script are discussed in the software section.
		
	Figure \ref{fig:rbodiagram} summarizes the power control hierarchy and local area network at RBO. We accomplish power management and distribution at RBO with an APC 2200 XL Smart-Uninterruptible Power Supply (UPS) and a TrippLite MH30NET Power Distribution Unit (PDU). A single municipal power line provides 120 V power to the UPS, which, in turn, powers all other devices necessary for remote/autonomous observatory operation. As the schematic shows, the UPS provides power to the TCS PC, MDC, dome/telescope/mirror door/focus motors, PDU, weather stations, science devices, radio, ethernet hub, and dome camera and lights. This represents a total maximum power load of about 1035 W, requiring a UPS rated to at least 1.28 kVA. Appendix \ref{app:power} is a tabulation of estimated power requirements.  A UPS battery extension module increases the emergency power runtime to a total of fifteen minutes in the event of a municipal power loss. An emergency power shutdown takes approximately five minutes to complete with no human interaction necessary, but reinitializing operations once power is restored does require a human operator. The PDU distributes power to the various observatory devices --- TCS PC, MDC, dome lights, and science devices --- via sixteen 120 V receptacles which are independently controllable through a web-based interface on the PDU's  network card. The PDU is programmed to schedule power events, such as enabling or disabling power to all devices on a daily basis. In the event of a municipal line power outage, the UPS triggers a dome closure through a single TTL output that connects to the MDC and provides the TCS with a signal to close. Through an AP9617 network card, the UPS is configured to send emergency emails to alert necessary contacts of the power loss.
	
	There are two weather stations installed at RBO: an Aurora Eurotech Cloud Sensor III and a Davis Vantage Vue. Both weather stations are powered directly by the UPS to ensure continuous operation. The Aurora unit measures wetness, light, and the presence of clouds; it provides a signal through a TTL output that connects to the MDC and provides the TCS with a trigger for precipitation or overcast induced dome closure. In the event of a dome closure signaled by this line, image capture continues and affected images are manually removed during analysis. The Davis weather unit measures humidity and wind data, uploads a weather record to a public website\footnote{\url{http://www.weatherlink.com/user/uwyorbo/}}. Together, the weather stations help ensure full observatory weather protection and network loss protection. Both stations were calibrated experimentally to establish the cautionary thresholds (40 km hr\textsuperscript{-1} windspeeds or 80\% humidity).

	Scientific devices on the telescope include two cameras, an Andor Apogee \textit{Alta F16} 4096$\times$4096 with 9 $\mu$m pixels and an approximately  25\arcmin$\times$25\arcmin\ field of view (FOV) and an Andor Apogee \textit{Aspen CG47} 1024$\times$1024 with 13 $\mu$m pixels and a 9\arcmin$\times$ 9\arcmin\ FOV, given the 5.06 m focal length of the telescope (plate scale of 0.04 \arcsec $\mu$m\textsuperscript{-1}). Either camera may be mounted behind a modified DFM 8-slot filter wheel, typically filled with 2 inch $\times$ 2 inch Bessel filters (\textit{UBVR$_C$I$_C$}), H-$\alpha$, \ion{O}{3}, and an empty slot. As seen in the schematic, the TCS PC controls these science devices over USB (cameras) and serial (filter wheel) connections. Given the approximate 6 m distance between the telescope and TCS PC, the cameras use an active USB extension cable to ensure reliable connectivity. 

	Dome lights and a dome camera provide additional environmental monitoring at RBO. The PDU allows activation of the dome lights for use by a remote observer. The dome camera is a D-Link DCS-933L ethernet home security camera with IR capabilities and an audio module mounted on the east interior dome-ring wall. The dome lights and dome camera only provide a passive means to monitor the observatory and are not utilized during automated observations.
 
	\subsection{Operation Software}
	ASCOM provides a consistent interface for all necessary observatory devices. There are many advantages to the ASCOM environment, but primarily automation benefits from the ease of interfacing with other software packages and the ability to script additional functionality into the system. ASCOM requires a Windows operating system. The upgrades performed by DFM to the TCS software provided the additional ASCOM functionality required to integrate devices and perform autonomous operation. Specifically, DFM added ASCOM functionality to dome control, dome shutter status, and telescope parking in the TCS software and created an ASCOM accessible mirror door control (as there exist no ASCOM objects or methods for mirror doors). These upgrades allowed us to develop full automation using a single commerical package and two in-house Python scripts\footnote{These scripts are available freely upon request from \email{physics@uwyo.edu}.}.
	
	MaxIm DL\footnote{MaxIm products available at \url{www.cyanogen.com/maxim\_main.php}} version 5 provides camera control and observatory automation via Visual Basic ASCOM scripting. This software can control a majority of commercial cameras. MaxIm also allows multiple devices to be controlled simultaneously, including cameras, focusers, and filter wheels. This program was chosen because it was already in use at RBO. As the rest of the observatory is ASCOM compliant, we use a combination of Python and Visual Basic scripts to prepare an automated observing run solely through MaxIm. Another popular automation software suite is CCDWare's Navigator and Autopilot \footnote{CCDWare products available at \url{www.ccdware.com}}. A major appeal of CCDWare's products is the user friendly environment. The CCDWare package has also been used successfully with RBO. With all of the hardware interfaced via ASCOM, it is possible to in-house script the entire automation process, removing the need to use additional commercial packages. We are working towards minimizing commercial software use --- Appendix \ref{app:code} contains a minimal example of controlling multiple ASCOM devices, which is the starting point for this project.

	A local Python operations script, called by the commercial package, controls mirror door and dome shutter operations. Mirror doors are always closed during operation of the dome shutter during startup and shutdown to protect telescope optics. As mirror doors are not a standard ASCOM object, there is no native method to control them. DFM's upgrades allow ASCOM access to the mirror doors through the generic telescope class' method \textit{Action} which our script invokes. 
	
	The second Python script is a weather monitoring script that parses the weather data from the Davis station and commands a temporary dome closure, using the same functionality of the operations script, if wind or humidity exceeds thresholds. The weather monitoring script uses Python module BeautifulSoup\footnote{More information about this module is available at \url{http://www.crummy.com/software/BeautifulSoup/}} to extract RBO's weather data from the website and then analyze it to determine if conditions are suitable for operation. The Davis station occasionally fails to upload weather data at regular intervals, and in this event, the conditions at the Laramie Regional Airport are analyzed instead\footnote{\url{http://weather.gov/82070}}. When analyzing airport data, we require the script weather thresholds be  10\% more conservative to account for the distance to the local station and reporting delays. If neither website can be reached, or if weather data timestamps are older than ten minutes, our script interprets this as a network failure and triggers a temporary dome closure until connections are reestablished.
	
	WinSCP\footnote{Information about WinSCP is available at: http://winscp.net/eng/docs/}, operating in batch-mode, transfers science images in real-time to the Physics \& Astronomy computer cluster archive. WinSCP automatically launches as a task after the user logs on to the TCS PC. WinSCP compares a directory at the observatory to a remote campus directory periodically and uploads any files not present on the remote machine. Copies of all data are stored on a hard drive at the observatory and a drive on campus.

\section{Nightly Operations}
\begin{figure}
	\begin{center}
		\includegraphics[width=.8\linewidth]{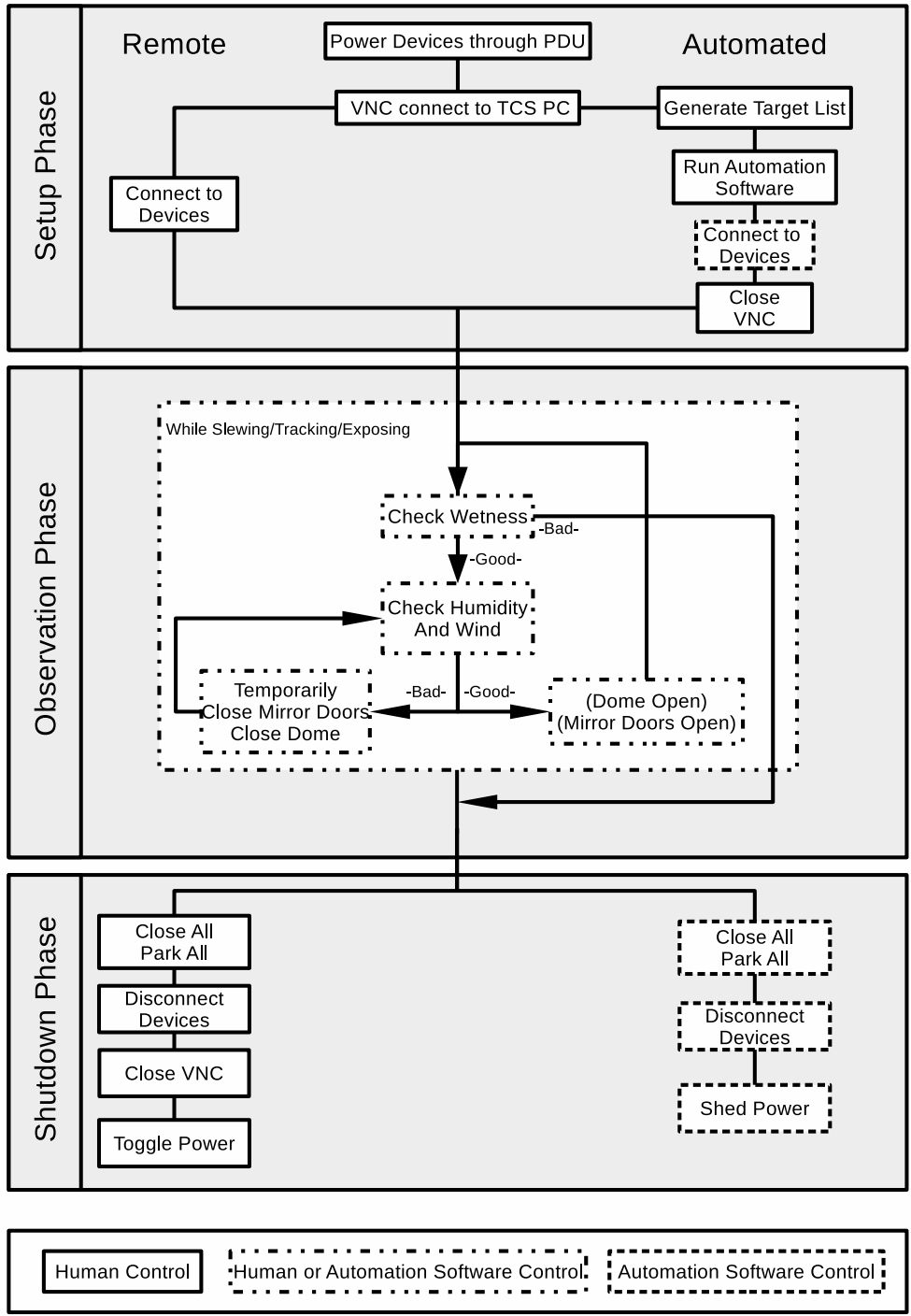}
		\caption{\label{fig:night}A simplified flowchart of an observation night at RBO in either remote- or automated-mode. Line styles given in the key denote observation control mode.}
	\end{center}
\end{figure}
There are many similarities between remote and automated observations. Figure \ref{fig:night} is a flowchart describing the sequence of events during full night's operation at RBO, with both remote and automated modes shown. Both modes involve three basic tasks required of any astronomical observatory --- setup, observe, and shutdown --- which are denoted by the three vertically arranged labelled grey regions in the Figure. Boxes in the left and right columns distinguish actions for remote and automated modes, respectively. Boxes in the middle column signify actions that occur in both remote and automated modes. In each mode there are human-controlled actions and software-controlled actions, designated by boxes with solid and dashed borders, respectively. Boxes with dash-dot border indicate actions that are preformed by a human in remote mode and by automation software in automated mode. Arrows designate the continuous observatory monitoring loops. 

Any remote or automated operation begins by a user accessing the PDU and powering on observatory devices. The user then connects to the TCS PC via Virtual Network Computing (VNC) protocol, allowing them to interact with the TCS PC as if it were local. The VNC server configuration in use at RBO allows remote users to control the computer immediately on startup. 

Once the user is logged on to the TCS PC, the action path diverges, depending on observation mode. For a remote-mode observing run, the observer establishes software connections to the relevant devices through MaxIm and TCS software. The remote observer is then finished with setup. In contrast, for an automated run, the user must generate target lists for either the MaxIm or CCDWare automation software. The automation software optimizes the observation sequence based on airmass and twilight constraints. In general, target selection and optimization is a complex issue --- the methods used at RBO will be discussed in the following section. The automation software then automatically establishes device connections, and the user closes the VNC connection, concluding the automation setup phase.

The observation phase for both modes is identical to classical observing with respect to the actions performed. Telescope slewing and tracking, camera temperature regulation and exposures, focus corrections, filter wheel selection, and dome movement are all completed. A remote-mode observer achieves this exactly as a classical observer, via direct interface with the TCS software and MaxIm. Automated-mode operations utilize the automation software to issue commands to control devices, as mediated by ASCOM. A notable exception to the automated solutions currently in use at RBO is the lack of automated focus adjustment --- at the time of writing this feature has not been implimented.

Throughout nightly observations, in either mode, two weather stations provide weather data to ensure observatory safety. In remote mode, the observer monitors wind and humidity and initiates dome shutter and mirror door closure, as necessary. Even in remote mode, the Aurora wetness sensor will close the dome in the event of active precipitation, necessitating a decision by the remote observer to reopen if conditions become safe. In automated mode, the Python weather monitoring script assesses humidity and wind data, initiating temporary closures until conditions are within limits for a consecutive five minutes. An automated night is declared over, without the possibility of re-opening, if the wetness sensor is tripped, given the lack of human assessment.

Observations end either when the sequence of targets is finished or when weather prohibits further operations. This begins the shutdown phase. A remote-mode observer closes mirror doors and dome shutters, parks the dome and telescope and disconnects all devices. Finally, they close the VNC connection and switch off all PDU-connected devices, protecting the devices from lightning, and thereby completing the remote-mode operations. Similarly, in automated-mode, the automation software runs the end-of-night sequence, including our Python operation script, and the observatory closes. The automation software disconnects from all devices. At morning twilight, the PDU powers off all connected devices according to a preprogrammed schedule, completing the automated operations sequence. 

\section{Exoplanets as a Science Driver at RBO}
Exoplanet transit photometry is the driving science case behind the RBO upgrades. Many small telescopes perform exoplanet transit photometric observations \citep{Pepper2007,Siverd2012}. Since completion of upgrades, both remote and automated operation modes have been employed to observe transit events. High-precision transit photometry demands tracking that prevents image motion across the detector. RBO's wide FOV combined with the the new absolute encoders leads to about 1\arcsec\ hr\textsuperscript{-1} drift on a single continuously tracked target. The absolute pointing is accurate to approximately 3\arcsec\ rms over the sky at less than 2 airmasses. These precisions allow RBO to perform high-precision transit photometry without a guide camera. In remote-mode operations, we find that the focus requires no adjustment during a typical week to accomplish successful transit photometry. Typical seeing of 3\arcsec\ for stellar point spread functions validates our lack of focus solution. In truth, focus is not important for transit observations and defocusing mitigates flat-fielding and fine guiding errors --- another strength of automated exoplanet photometry.

Scheduling photometric observations of exoplanets at RBO is accomplished with the TAPIR code \citep{Jensen2013} and operator decision making. This tool takes target lists with ephemerides, periods, and coordinates and produces a list of visible targets which are transiting within a time frame. On a given night, TAPIR typically produces no more than twenty targets. From these targets, a human operator decides the transit(s) to observe through knowledge of RBO's observational limits, weather forecasts, and fullness of transit event. In practice, this means scheduling and optimization of the night's observations is controlled by the time-domain targets. Logistics of scheduling and optimizing observations will vary between projects.

We observed a transit of TrES-3b (an $\sim$ 12 mag star in R) with the Andor Apogee \textit{Aspen CG47} camera in the Bessel \textit{R$_C$} filter at RBO on 2016-05-04 UTC. Observations started at 4:40 UT and lasted approximately 3.5 hours, during which we obtained 31 frames. The exposure time for all frames was 200 s and at a camera temperature of -30\degr C throughout the entire observation. We used six stars as differential photometric standards. The field was densely populated and standards were chosen based on signal strength and lack of overexposure. The only comparison existing in a catalog was USNO-B1.0 1275-00332540. Sky conditions were photometric at first and became less amenable to observing towards the end of the the transit. This resulted in a typical seeing of approximately 3\arcsec\ FWHM over the 1.10--1.55 airmass range of observations. Images were bias, dark, and flat corrected using sequences of 50 bias frames, 25 dark frames taken at the same camera temperature, and 30 flat-field frames taken manually on twilight sky that night. Photometry was preformed on the target star and standards with the AstroImageJ package \citep{Collins2016} using apertures of 10\arcsec\ radius on all stars. Average frames contained $2.4\times10^{6}$ photons for the target star with formal uncertainties of 1.0 mmag.

\begin{figure}
	\begin{center}
		\includegraphics[width=.9\linewidth]{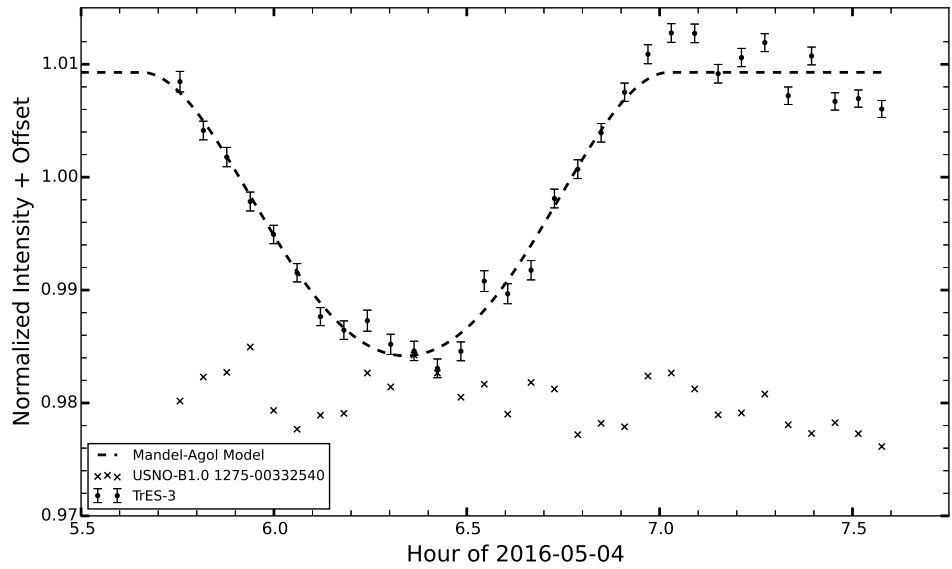}
		\caption{\label{fig:transit}The light curve of the transiting system TrES-3, as recorded at RBO starting at 2016-05-03 05:45 UTC. Filled circles with error bars denote observations of the TrES-3b transit from a 200 s exposure. A theoretical curve obtained using the Mandel-Agol model is overlain. The crosses are USNO-B1.0 1275-00332540 comparison star observations exemplifying the measured sky variability during the transit.}
	\end{center}
\end{figure}
Figure \ref{fig:transit} shows TrES-3's light curve during a planetary transit, exemplifying the photometric performance of RBO. Solid circles with error bars are the photometric data from a 200 s exposure. To demonstrate the precision of our data, the Figure also shows a \citet{Mandel2012} transit light curve model as a dashed line. The light curve model was generated with orbital parameters from published values \citep{Parviainen2016} and created by the BATMAN code \citep{Kriedberg2015}. The good agreement between our data and the synthetic light curve illustrates the potential for RBO to conduct high-precision photometric campaigns and detect planetary transits at the few mmag level. Most points fall on the expected curve within 1$\sigma$, with more variability occurring at the end of the transit due to the development of non-photometric observing conditions.

\section{Conclusion}
Hardware and software upgrades are now completed that have made remote and automated operations possible at the Red Buttes Observatory. The automated observatory now allows a year-round duty cycle with much greater flexibility of operations, dramatically increasing the number of observed nights. This has spurred the inception of multiple projects covering a range of research and pedagogical interests. Both graduate and undergraduate students have acquired expertise in observatory operations, automation, and data acquisition and analysis during the upgrade process. Undergraduate students routinely conduct transit follow-up photometry and participate in an authentic research program. Additionally, laboratory classes now use RBO to educate and expose undergraduates to astronomical techniques early in their college careers. This modernization has invigorated the legacy observatory, increasing RBO's capability to carry out its teaching and research mission.

\section*{Acknowledgments}
This project was supported by Wyoming NASA Space Grant Consortium, NASA Grant \#NNX15AI08H, Wyoming EPSCoR, and the Wyoming Research Scholarship Program. We are grateful to UW undergraduate Aman Kar for his assistance in observing targets. Additionally, communication with Scott Mecca and Chris Rice at the outset of the project were invaluable to our success. 

\appendix
\section{Estimations of Power Usage\label{app:power}}
We present an estimation of the power usage at RBO in a worst-case scenario. That is, this estimation assumes all devices are simultaneously operating at maximum load. This is never the case, as some actions cannot be performed at the same time, but it provides an upper limit. The estimations include devices not discussed in the text as they are not integral to automated operations, but are, nevertheless, part of the power load at RBO. The load of a power supply is often given in units of volt-amps, while device loads are in watts. The conversion volt-amp to watts is known as the power factor: (PF) = $\dfrac{\text{W}}{\text{VA}}$. The RBO maximum power load of 1035 Watts requires a UPS rated to 1035/PF; about 1280 VA for our UPS' power factor of 0.84.
\begin{table}[H]
	\begin{center}
		\begin{tabular}{l|r|r}
		Device 	& Quantity 	& Load [W] \\\hline
		PC	& 2		& 565\\
		Dome Motor	&	& 186\\
		Dome Shutter	&	& 100\\
		Camera		&	& 75\\
		Mirror Doors	&	& 41\\
		Telescope	&	& 33\\
		Monitor		& 2	& 35\\\hline\hline
			& Total Power Load& \textbf{1035}
		\end{tabular}
	\end{center}
\end{table}

\section{ASCOM Control with Python}\label{app:code}
The ASCOM website gives examples of scripting in many coding languages, but not Python. As Python is such a ubiquitous language in the physical sciences, we present a minimal working example of interfacing with observatory hardware as translated from the ASCOM guide for other languages. In this example, we invoke our particular driver (TCSGalilHybrid), which is specific to our hardware installation. As ASCOM is only available on Windows, this program will also only run on Windows.
\footnotesize
\begin{verbatim}
#Import the Python module that negotiates Windows COM communications
import win32com.client as com

#Create telescope, dome, and camera objects
driver = 'TCSGalilHybrid' #the name of our specific driver

#Any ASCOM object (i.e. dome, camera, rotator, focuser) is initialized as
# com.Dispatch('DriverName.Object')
tele = com.Dispatch(driver+'.Telescope')
dome = com.Dispatch(driver'.Dome')

#We choose to use the MaxIm controls becuase they are native to the camera
#hardware and avoid unnecessary 3rd party control
camera = com.Dispatch('MaxIm.CCDCamera')

#Allow commands to be sent to the devices
tele.Connected = True
dome.Connected = True

#Prepare dome for observation
dome.OpenShutters()
dome.Slaved = True #Have the dome follow the telescope's azimuth

#Prepare telescope for observation
tele.Tracking = True
tele.TrackingRate = 15 #arcsec/sec

#Prepare camera for observation. The camera commands are for MaxIm as opposed
#to ASCOM and so do not follow ASCOM style 
camera.LinkEnabled = True 
camera.CoolerOn = True
camera.TemperatureSetpoint = -30. #Degrees Celsius

#Slew the telescope to target location
#The Async issues the command without pause
tele.SlewToCoordinatesAsync(101.2875, -12.72) #Dec. hours, dec. deg.

#Take Exposures
#A 25 second exposure, the "1" states that this is a light frame
camera.Expose(25, 1)
#Wait until exposure has finished integrating
while not camera.ImageReady:
    pass
camera.SaveImage('E:\Data\Sirius.fit')

#Here we would return the telescope to zenith, free the dome,
#and warm the camera

#Disconnect the COM connections
camera.Connected = False
tele.Connected = False
dome.Connected = False
#Remove the objects so as to prevent COM issues on closure
del camera, tele, dome

\end{verbatim}\normalsize
Our complete control program utilizes many other modules for other functionality such as coordinate precession and managing target lists. This code is too lengthy to put as an appendix and as such is available upon individual request. The full functionality of each object is available on the ASCOM webpage present in the body. The scripting guide for a MaxIm controlled camera is available at \url{http://www.cyanogen.com/help/maximdl/MaxIm-DL.htm} . 
\bibliography{ms}

\begin{thebibliography}{}
\expandafter\ifx\csname natexlab\endcsname\relax\def\natexlab#1{#1}\fi

\bibitem[{{Collins} {et~al.}(2016){Collins}, {Kielkopf}, \&
  {Stassun}}]{Collins2016}
{Collins}, K.~A., {Kielkopf}, J.~F., \& {Stassun}, K.~G. 2016, ArXiv e-prints,
  arXiv:1601.02622

\bibitem[{{Jensen}(2013)}]{Jensen2013}
{Jensen}, E. 2013, {Tapir: A web interface for transit/eclipse observability},
  Astrophysics Source Code Library, ascl:1306.007

\bibitem[{{Kiminki} {et~al.}(2015){Kiminki}, {Kobulnicky}, {Vargas
  {\'A}lvarez}, {Alexander}, \& {Lundquist}}]{Kiminki2015}
{Kiminki}, D.~C., {Kobulnicky}, H.~A., {Vargas {\'A}lvarez}, C.~A.,
  {Alexander}, M.~J., \& {Lundquist}, M.~J. 2015, \apj, 811, 85

\bibitem[{{Kreidberg}(2015)}]{Kriedberg2015}
{Kreidberg}, L. 2015, \pasp, 127, 1161

\bibitem[{{Mandel} \& {Agol}(2002)}]{Mandel2012}
{Mandel}, K., \& {Agol}, E. 2002, \apjl, 580, L171

\bibitem[{{Monson} \& {Pierce}(2009)}]{Monson2009}
{Monson}, A.~J., \& {Pierce}, M.~J. 2009, \pasp, 121, 728

\bibitem[{{Parviainen} {et~al.}(2016){Parviainen}, {Pall{\'e}}, {Nortmann},
  {Nowak}, {Iro}, {Murgas}, \& {Aigrain}}]{Parviainen2016}
{Parviainen}, H., {Pall{\'e}}, E., {Nortmann}, L., {et~al.} 2016, \aap, 585,
  A114

\bibitem[{{Pepper} {et~al.}(2007){Pepper}, {Pogge}, {DePoy}, {Marshall},
  {Stanek}, {Stutz}, {Poindexter}, {Siverd}, {O'Brien}, {Trueblood}, \&
  {Trueblood}}]{Pepper2007}
{Pepper}, J., {Pogge}, R.~W., {DePoy}, D.~L., {et~al.} 2007, \pasp, 119, 923

\bibitem[{{Rodgers} {et~al.}(2006)}]{Rodgers2006}
{Rodgers}, C., {et~al.} 2006, GRB Coordinates Network, 4830

\bibitem[{{Siverd} {et~al.}(2012){Siverd}, {Beatty}, {Pepper}, {Eastman},
  {Collins}, {Bieryla}, {Latham}, {Buchhave}, {Jensen}, {Crepp}, {Street},
  {Stassun}, {Gaudi}, {Berlind}, {Calkins}, {DePoy}, {Esquerdo}, {Fulton},
  {F{\H u}r{\'e}sz}, {Geary}, {Gould}, {Hebb}, {Kielkopf}, {Marshall}, {Pogge},
  {Stanek}, {Stefanik}, {Szentgyorgyi}, {Trueblood}, {Trueblood}, {Stutz}, \&
  {van Saders}}]{Siverd2012}
{Siverd}, R.~J., {Beatty}, T.~G., {Pepper}, J., {et~al.} 2012, \apj, 761, 123

\end{thebibliography}

\end{document}